\documentclass[conference]{IEEEtran}
\usepackage{graphicx}
\usepackage{hyperref}
\usepackage{multirow}

\title{Robust Tracking and Behavioral Modeling of Movements of
  Biological Collectives from Ordinary Video Recordings}

\author{
\IEEEauthorblockN{Hiroki Sayama\IEEEauthorrefmark{1}\IEEEauthorrefmark{2}, Farnaz Zamani Esfahlani\IEEEauthorrefmark{1},
Ali Jazayeri\IEEEauthorrefmark{3} and J.\ Scott Turner\IEEEauthorrefmark{4}}
\IEEEauthorblockA{\IEEEauthorrefmark{1}Center for Collective Dynamics of Complex Systems /
Department of Systems Science and Industrial Engineering\\
Binghamton University, State University of New York, Binghamton, NY 13902-6000}
\IEEEauthorblockA{\IEEEauthorrefmark{2}Faculty of Commerce, Waseda University, Shinjuku, Tokyo 169-8050, Japan}
\IEEEauthorblockA{\IEEEauthorrefmark{3}Department of Information Science, College of Computing \& Informatics, Drexel University, 
Philadelphia, PA 19104}
\IEEEauthorblockA{\IEEEauthorrefmark{4}Department of Biology, SUNY College of Environmental Science and Forestry, Syracuse, NY 13210}
Email: sayama@binghamton.edu
}

\begin{document}
\maketitle

\begin{abstract}
We propose a novel computational method to extract information about
interactions among individuals with different behavioral states in a
biological collective from ordinary video recordings. Assuming that
individuals are acting as finite state machines, our method first
detects discrete behavioral states of those individuals and then
constructs a model of their state transitions, taking into account the
positions and states of other individuals in the vicinity. We have
tested the proposed method through applications to two real-world
biological collectives: termites in an experimental setting and human
pedestrians in a university campus. For each application, a robust
tracking system was developed in-house, utilizing interactive human
intervention (for termite tracking) or online agent-based simulation
(for pedestrian tracking). In both cases, significant interactions
were detected between nearby individuals with different states,
demonstrating the effectiveness of the proposed method.
\end{abstract}

\section{Introduction}

Motion tracking and behavioral modeling of biological collectives,
such as bird flocks, fish schools, insect swarms, and pedestrian
crowds, have been the subject of active research in behavioral ecology
\cite{kabra2013jaaba,perez2014idtracker,dell2014,vela2015individual}
and artificial life
\cite{kunz2003artificial,sayama2009swarm,hamann2014analysis}. This is
arguably one of the most challenging tasks in complex systems science,
in which researchers need to elucidate unknown microscopic rules that are
responsible for the observed or desired macroscopic emergent
behaviors.

Previous literature on this topic was mostly focused on
modeling individual behaviors in isolation or in response to other
individuals in the vicinity through simple kinetic interaction
rules. A commonly adopted assumption is that the same set of
behavioral rules applies to all individuals homogeneously over space
and time, while not much attention was paid to modeling and analyzing
multiple distinct behavioral states and their interactions in the
collective. However, considering the possibility of multiple
behavioral states and their interactions would make the models closer
to real biological collectives, where individuals are not acting like
non-living particles but can switch from one behavior to another
dynamically \cite{sayama14four}.

The objective of the present study is to (a) propose and evaluate a
new, generalizable computational method to detect and model
multiple distinct behavioral states and their interactions among
individuals from their externally observable spatio-temporal
trajectories, and to (b) develop robust motion tracking systems that
can generate trajectory data from ordinary, low-resolution video
recordings of real-world biological collectives. In what follows, we
first describe the foundational theoretical method of behavioral
modeling, and then discuss two versions of robust tracking systems
that were developed and tested through applications with two
real-world examples of biological collectives: termites and human
pedestrians.

\section{Method for Behavioral Modeling}

In this section, we describe theoretical details of the proposed
behavioral modeling method. We construct a stochastic model of
behavioral transitions of individuals that depend on both the internal
state of the individual and the external environmental context,
including presence and states of other individuals nearby.

The proposed method analyzes the following observational data for each
individual $i$ in population $S$ ($i \in S$; $S$ can vary over time):
\begin{itemize}
\item $\vec{x}_i (t)$ : Position of individual $i$ at time $t$
\item $E_i(t)$ : Local environmental condition for individual $i$ at time $t$
\end{itemize}
Here, $E_i(t)$ may include other individuals ($\{\vec{x}_{j\neq
  i}(t)\}$) and other environmental constraints (e.g., obstacles
placed in the space, etc.) that are present in the vicinity of
individual $i$. The spatial/temporal radii that determine the ``local
environment'' are important parameters, which will depend on the
specific nature of the collective under investigation. Note that these
two pieces of data, $\vec{x}_i (t)$ and $E_i(t)$, are purely
observational, described from the viewpoint of an
observer/experimenter outside the collective.

Our objective is to computationally convert these observational data
into a more dynamical, rule-based description of the collective
behavior, with possible interactions among the individuals also
adequately captured. To facilitate this challenging task, we adopt a
{\em finite state machine} assumption, i.e., that the states of
individuals in the collective are discrete and finite, and that they
can change dynamically according to the input coming from their local
environment \cite{sayama14four}.

The proposed method consists of the following two steps:

\subsection{Step 1. Identifying discrete behavioral states}

The objective of this step is to construct the following behavioral
state function for each individual $i$ at time $t$:
\begin{equation}
b_i(t) = \mathcal{L}(\vec{x}_i, E_i) (t) \quad \quad b_i(t) \in B
\end{equation}
Here, $B$ is a set of discrete behavioral states, and the operator
$\mathcal{L}$ applies predefined mathematical operations or
computational algorithms to the time series data $\vec{x}_i$ and $E_i$
(not necessarily just their localized values at time $t$) and produces
a behavioral state function $b_i$. Examples of $\mathcal{L}$ include
clustering analyses applied to the frequency components of a short
segment of $\vec{x}_i(t)$ around $t$ (used in the first application),
and to the velocity vectors of individuals at $t$ (used in the second
application). Specific choices of $\mathcal{L}$ will depend on the
properties of the collective being analyzed.

We also note that, once $b_i(t)$ has been calculated for $i$ and $t$,
it can be embedded into $E_i(t)$ and second-order (and higher-order)
behavioral state functions could be calculated recursively using the
updated $E_i(t)$. While this is interesting from a ``theory of mind''
viewpoint, we did not do such recursive, higher-order behavioral state
identification in the present study. However, the first-order
behavioral states $b_i(t)$ were embedded in $E_i(t)$ for the following
``Building state transition rules'' step, in order to detect local
interactions of individuals that were specific to their states.

\subsection{Step 2. Building state transition rules}

The objective of this step is to construct rules of state transitions
that were observed in $b_i(t)$, using $\vec{x}_i(t)$ and
$E_i(t)$. Because of the large amount of stochasticity typically
present in the observational data, it is reasonable to assume that the
transition rules are described in the form of state transition
probabilities. This can be written mathematically as follows:
\begin{equation}
\vec{p}_i(t+\Delta t) \approx \mathcal{T}(\vec{x}_i(t), b_i(t),
E_i(t)) \label{eq:transition}
\end{equation}
Here, $\vec{p}_i(t+\Delta t)$ is the probability vector with dimension
$|B|$, which represents the probability distribution of the next state
of individual $i$ after a short period of time $\Delta t$, and
$\mathcal{T}$ represents the state transition rules that receive the
current position, behavioral state, and local environment of individual $i$ and
generate $\vec{p}_i(t+\Delta t)$. Unlike $\mathcal{L}$ in the first
step, $\mathcal{T}$ is an unknown mechanism that is to be estimated in
this second step, and some heuristic assumptions would be needed to
make this estimation task manageable.

In the present study, we assumed that the state transitions would not
depend on the position of the individual (this eliminates
$\vec{x}_i(t)$ in Eq.~(\ref{eq:transition})), and that $\mathcal{T}$
would simply be a matrix uniquely defined for each behavioral state
(which we call $T_{b_i(t)}$ hereafter) that generates
$\vec{p}_i(t+\Delta t)$ when multiplied with an environment vector
$E_i(t) = \vec{e}_i(t)$. The contents of $\vec{e}_i(t)$ will depend on
specific applications. With these assumptions,
Eq.~(\ref{eq:transition}) is simplified as follows:
\begin{equation}
\vec{p}_i(t+\Delta t) \approx T_{b_i(t)} \vec{e}_i(t)
\end{equation}
This simplification was partly based on our earlier work on
morphogenetic swarm chemistry \cite{sayama14four}. With this
simplification, the estimation of $\mathcal{T}$ becomes a simple least
squares estimation of $T_{b_i(t)}$ for each $b_i(t)$, where
$\vec{p}_i(t+\Delta t)$ and $\vec{e}_i(t)$ are given in the
observational data.

Overall, the final result of this step is given as a $|B| \times |B|
\times m$ tensor $\hat{T} = \left(T_1 \; T_2 \; \ldots \;
T_{|B|}\right)$, where $m$ is the number of elements in the
environment vector $\vec{e}_i(t)$.

\section{Application I: Detecting Interactions Among Termites}

\begin{figure}
\centering
\includegraphics[width=0.9\columnwidth]{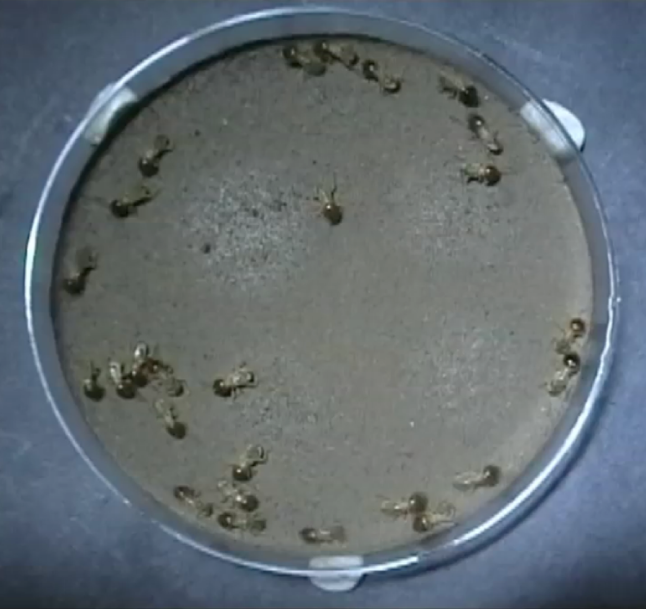}
\caption{A swarm of 26 individuals of termite species {\em Macrotermes
    michaelseni} moving in a Petri dish.}
\label{termites}
\end{figure}

We first tested the proposed method through an application to a
small-sized population of termites in a well-controlled experimental
setting in a laboratory. The objective of this task was to detect
interactions among a fixed number (26, specifically) of individuals of
termite species {\em Macrotermes michaelseni} moving in a Petri dish
(Fig.~\ref{termites}).

\begin{figure}
\centering
\includegraphics[width=\columnwidth]{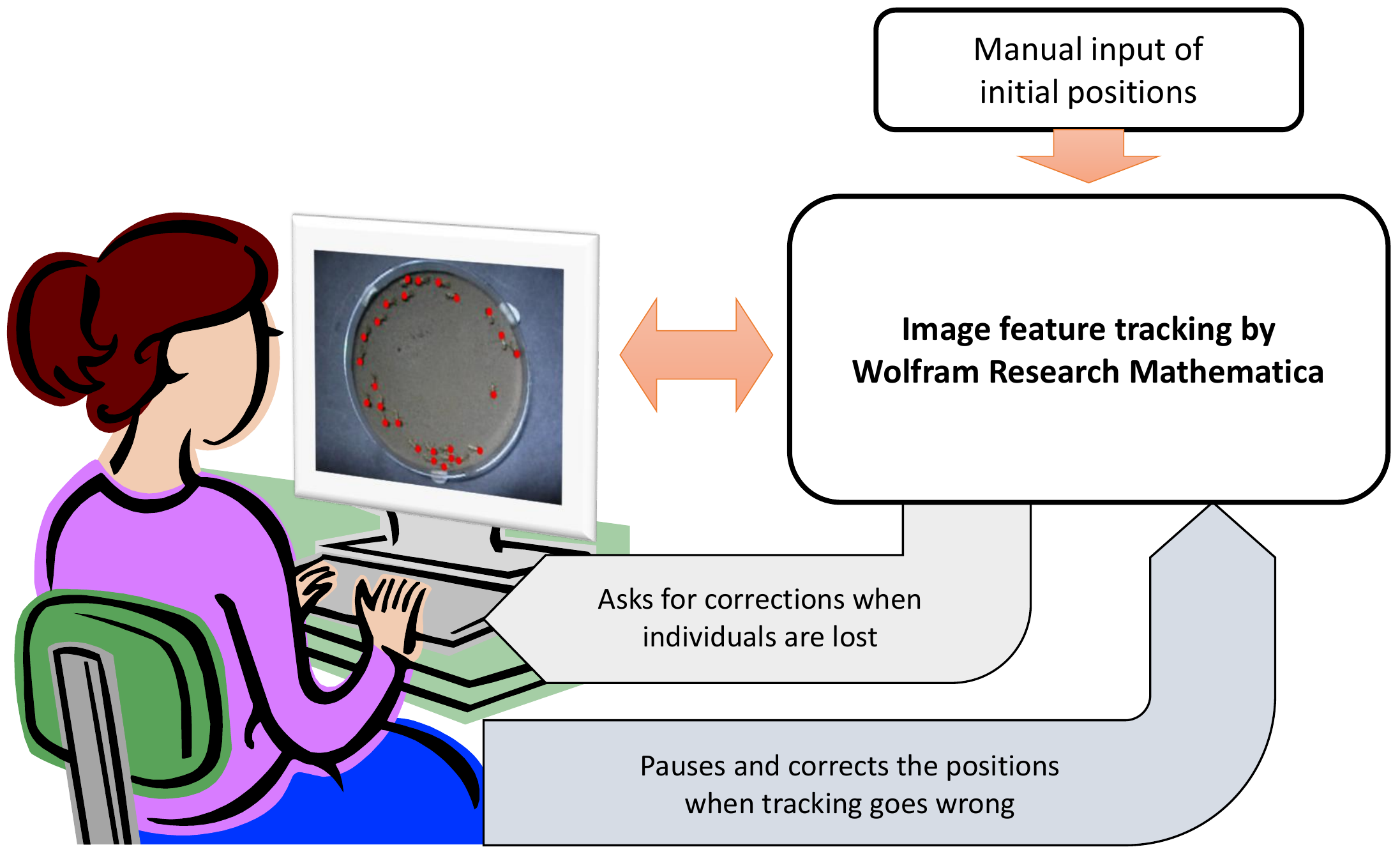}
\caption{Interactive, semi-automated tracking system for the first
  application (tracking termites) developed using Wolfram Research
  Mathematica's image feature tracking functions.}
\label{interactive-system}
\end{figure}

In order to detect and track the positions of individual termites
robustly from a low-resolution video recording, we developed an
interactive, semi-automated tracking system using Wolfram Research
Mathematica's image feature tracking functions
(Fig.~\ref{interactive-system}) \cite{sayama2015coco}. A human user
manually enters the initial positions of individuals, and then
continuously monitors the automated feature tracking process conducted
by Mathematica. When tracking goes wrong, either the human user pauses
the system and corrects the positions of tracked individuals, or the
system pauses itself and asks the human user for corrections. This
hybrid approach integrating automated tracking by a computer and
interactive intervention by a human observer resulted in an effective,
low-cost, robust solution that can work with ordinary low-resolution
video recordings.

\begin{figure}
\centering
\includegraphics[width=0.8\columnwidth,height=0.77\columnwidth]{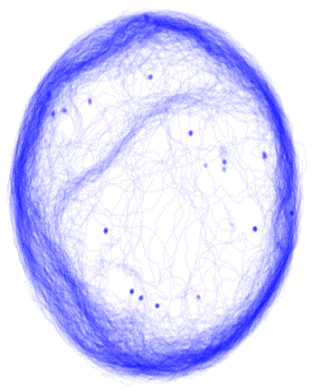}
\caption{Trajectories of termites detected using the interactive
  tracking system, visualized for all the individuals at once. Termite
  individuals were circling near the edge of the Petri dish most of
  the time, but some interesting emergent trails were also revealed.}
\label{termite-trajectories-all}
\end{figure}

\begin{figure}
\centering
\includegraphics[width=\columnwidth]{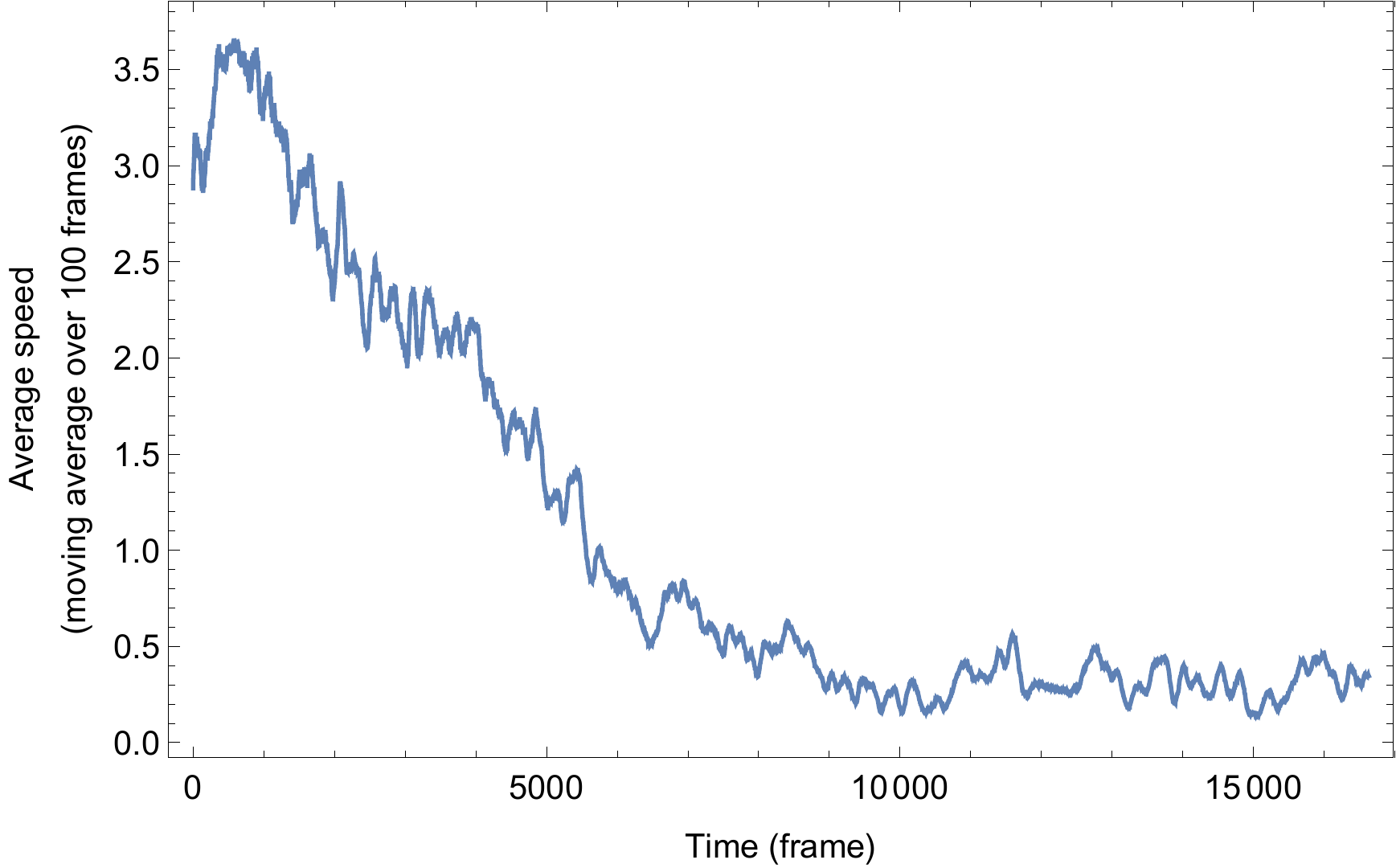}
\caption{Temporal change of activities of termites characterized by
  their average speed. Their activities decreased over time and
  eventually converged at a low stationary level.}
\label{termite-temporal-change}
\end{figure}

Using the developed tracking system, we converted the video recording
into a set of positional time series data, which already showed some
interesting emergent trails inside the Petri dish
(Fig.~\ref{termite-trajectories-all}). It was also observed that the
termites' activities characterized by the average speed decreased over
time, eventually converging at a low stationary level
(Fig.~\ref{termite-temporal-change}).

\begin{figure}
\centering
\includegraphics[width=\columnwidth]{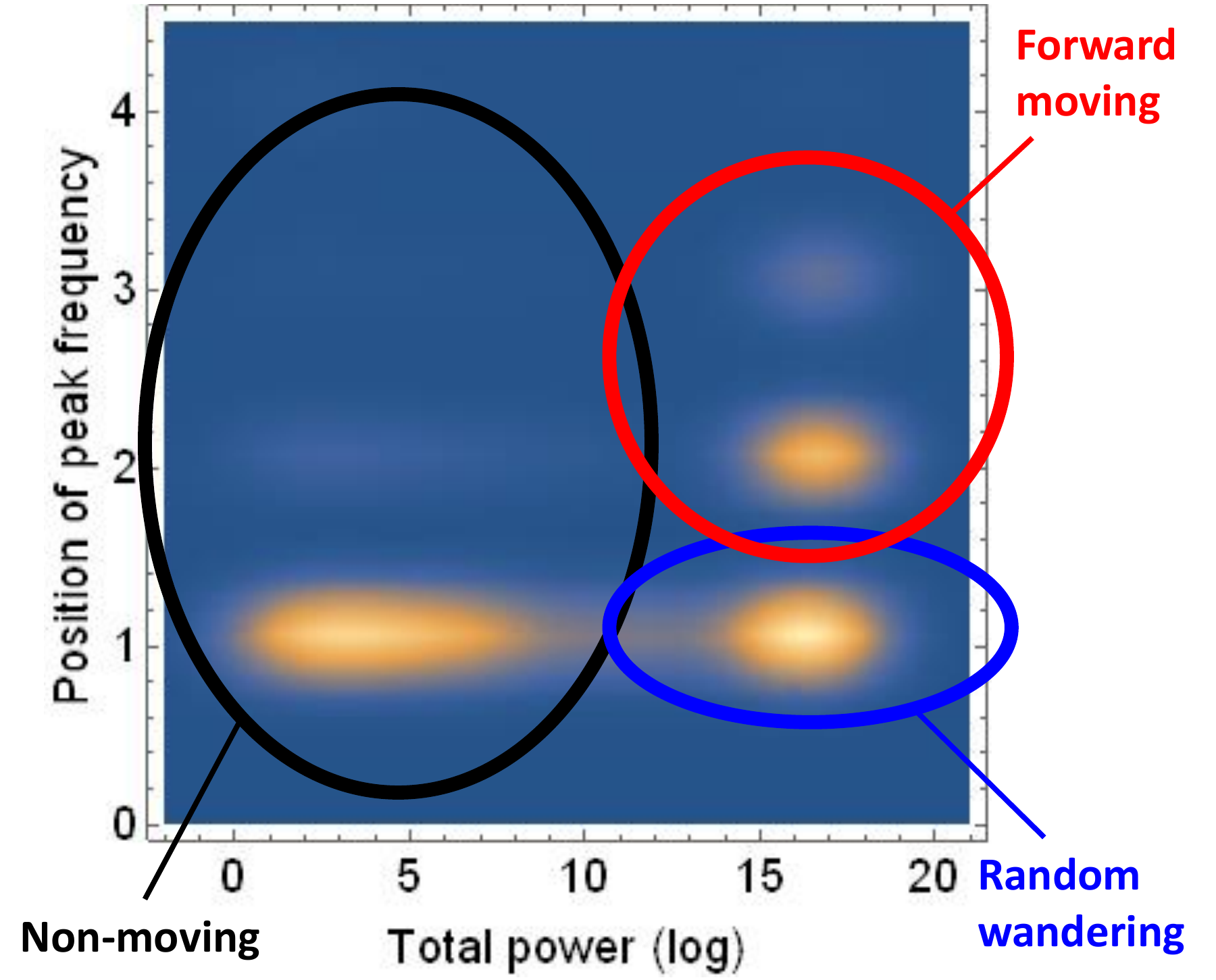}
\caption{Results of behavioral identification of termites. Three
  behaviors were identified: Non-moving (black), random wandering (blue)
  and forward moving (red).}
\label{termite-behaviors}
\end{figure}

\begin{figure*}
\centering
\includegraphics[width=\textwidth]{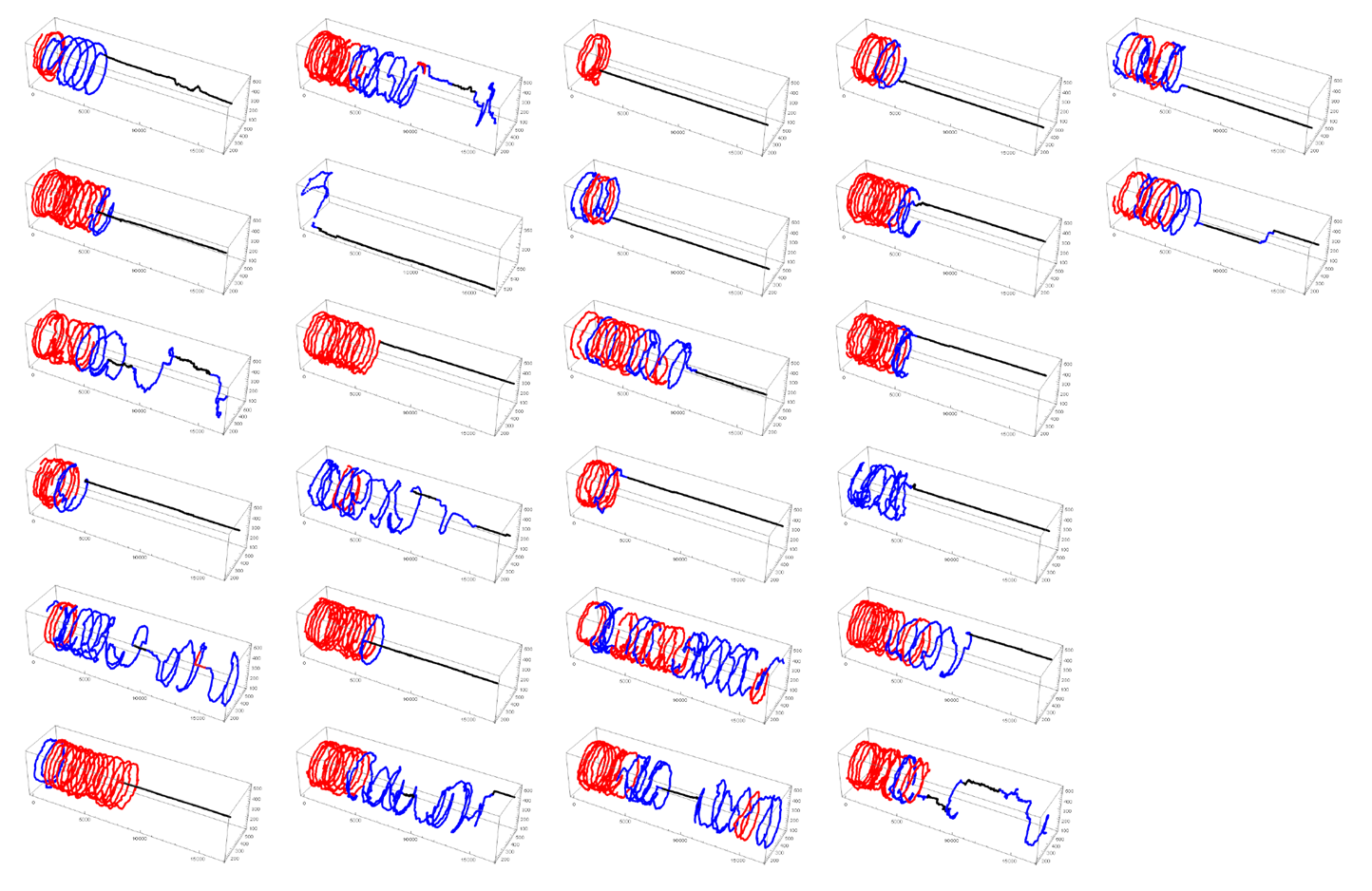}
\caption{Behavioral state functions ($b_i(t)$) of 26 termite
  individuals visualized on each individual trajectory. Time flows
  from left to right. Behavioral states are indicated by color
  (non-moving = black, random wandering = blue, and forward moving =
  red).}
\label{termite-trajectories}
\end{figure*}

For identification of discrete behavioral states, we calculated power
spectra of short segments (1,000 frames corresponding to about 30
seconds) of the trajectory for each individual around each time point,
and extracted the following two metrics from each spectrum: (1) total
power of ten lowest frequency components, and (2) peak
frequency. Clustering analysis applied to the two-dimensional
behavioral feature space created by these two metrics revealed several
distinct clusters, which we classified into three behaviors:
non-moving, random wandering, and forward moving
(Fig.~\ref{termite-behaviors}). The behavioral state functions
$b_i(t)$ were constructed using these identified behavioral states, as
shown in Fig.~\ref{termite-trajectories}.

\begin{figure*}
\centering \includegraphics[width=0.7\textwidth]{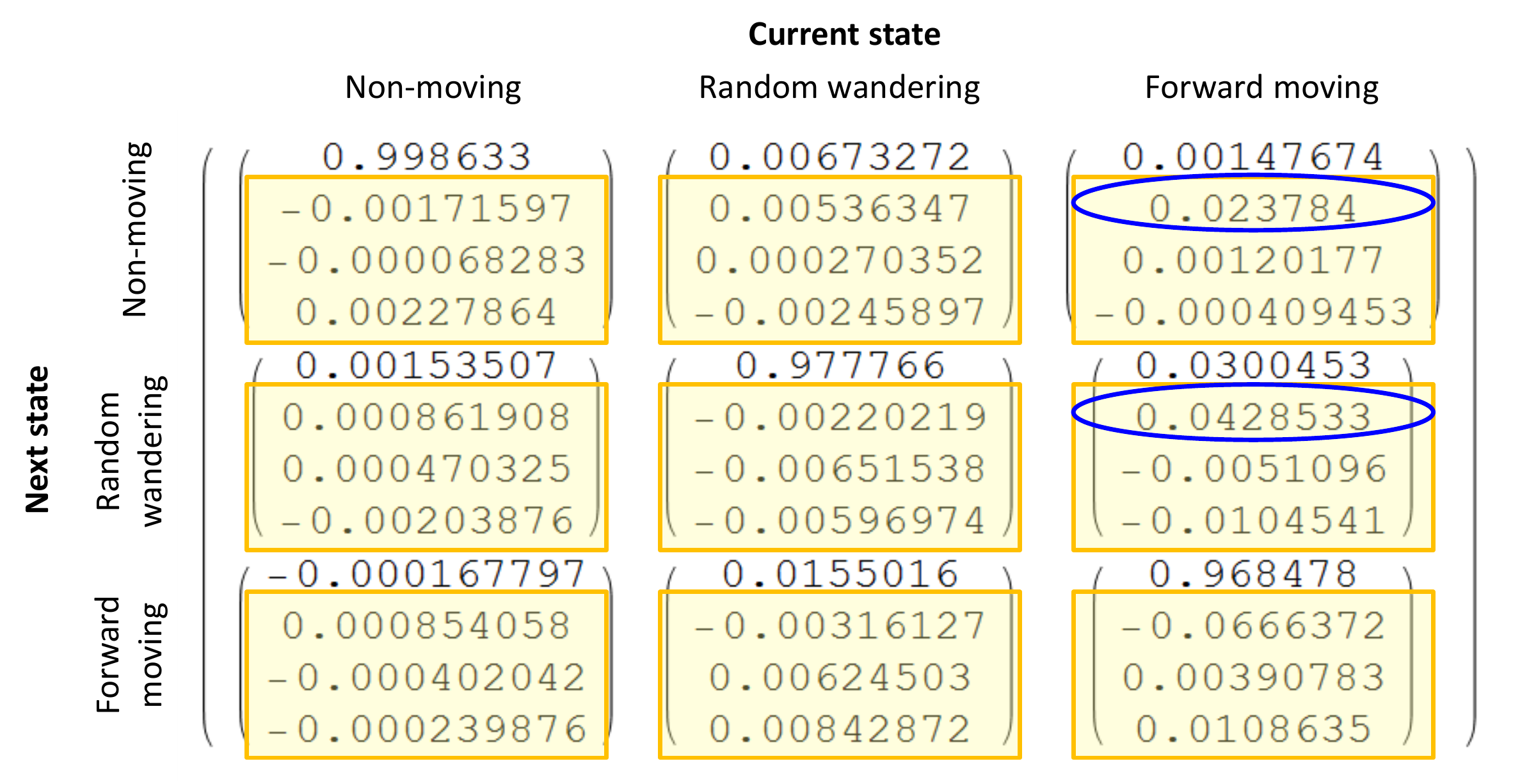}
\caption{State transition tensor $\hat{T}$ obtained from the
  trajectories of termites. The tensor was transposed to improve
  readability. The size of $\hat{T}$ is $3 \times 3 \times 4$. The
  numbers outside the yellow boxes represent inherent state transition
  probabilities from one behavioral state (shown at the top) to
  another (shown on the left). The numbers inside each yellow box
  represent changes of the state transition probabilities caused by
  the presence of another individual nearby with each of the three
  behavioral states (top: non-moving, middle: random wandering,
  bottom: forward moving, in each box).}
\label{termite-tensor}
\end{figure*}

To construct the state transition rules, we used the numbers of other
nearby individuals in each of the three behavioral states as the
environment vector $\vec{e}_i(t)$. The spatial radius of neighbor
detection was set to approximate the termite individuals' typical body
size, assuming that their individual-to-individual interactions were
largely contact-based. We also included a constant unity as the first
element of $\vec{e}_i(t)$ so as to be able to detect inherent state
transition probabilities that were independent of the influences of
nearby individuals. The state transition tensor $\hat{T}$ was
estimated using the standard least squares method for each
$T_{b_i(t)}$.

The results are shown in Fig.~\ref{termite-tensor}, where the
strengths of interactions among individuals in local vicinity were
successfully captured (marked with yellow boxes in
Fig.~\ref{termite-tensor}). Positive (negative) numbers indicate that
the presence of individuals with a particular state will increase
(decrease) the state transition probability. For example, the
particularly large values circled in blue in Fig.~\ref{termite-tensor}
show that, if a forward-moving termite individual faces another
individual that is not moving (we can tell this because these numbers
appeared in the top row of each yellow box), then the probabilities
for state transition to the non-moving or random wandering state will
increase significantly. Another example is 0.0108635 at the very
bottom to the right in the tensor; this means that, if a
forward-moving individual has another running mate nearby that is also
forward moving, then the probability of keeping the same state will
receive a boost by about 1\% (otherwise it would be 96.85\%, which is
given right above the bottom-right yellow box).

This result illustrates that the proposed method can provide
quantitative information about how individuals with various behavioral
states interact with each other, and how strong such interactions are,
in the biological collective being observed.

\section{Application II: Detecting Interactions Among Pedestrians}

\begin{figure}
\centering
\includegraphics[width=\columnwidth]{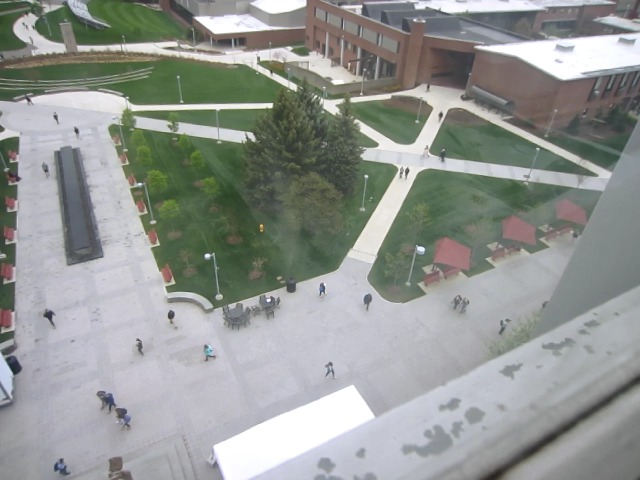}
\caption{A sample snapshot of birds-eye-view video recordings of
  pedestrians walking in the Binghamton University campus, taken from
  the ninth floor of the University's Library Tower.}
\label{birdseye-view}
\end{figure}

The second application is to track and model movements of human
pedestrians in a university campus. This is to test the applicability
of the proposed method to more complex, noisy collectives ``in the
wild'', with dynamically changing size. As the input data, we recorded
pedestrians walking in the central part of the Binghamton University
campus during a lunch break for one hour
(Fig.~\ref{birdseye-view}). The recording was conducted with the
Binghamton University IRB approval. During the recorded period of
time, a large number of pedestrians entered and moved out of the
recorded spatial area, making the size of $S$, the set of moving
individuals, dynamically changing. To make things more complicated,
there were several obstacles (e.g., trees, lampposts; see
Fig.~\ref{birdseye-view}) that would temporarily hide moving
individuals, making the tracking task very challenging.

\begin{figure}
\centering
\includegraphics[width=0.9\columnwidth]{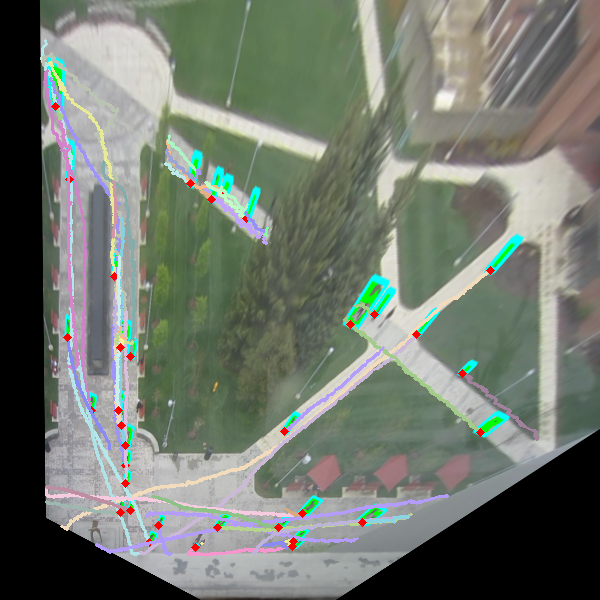}
\caption{Transformed top-view video, being analyzed by the hybrid
  tracking system that combines image processing and online
  agent-based simulation. Detected pedestrians are highlighted by cyan
  rectangles, with red dots at the bottom indicating the locations of
  simulated ``agents''. Their trajectories are shown in different
  colors. Black areas are the outside of the original video frame.}
\label{top-view}
\end{figure}

The positions of moving individual pedestrians were tracked using a
custom-made image processing program that was developed using OpenCV
\cite{bradski2008learning}. First, the background subtraction function
of OpenCV was utilized to detect objects that did not belong to the
static background. The perspective of the image was then transformed
from the birds-eye view to the top view, in which the positions and
velocities of the objects were calculated and tracked
(Fig.~\ref{top-view}).

\begin{figure}
\centering
\includegraphics[width=0.9\columnwidth]{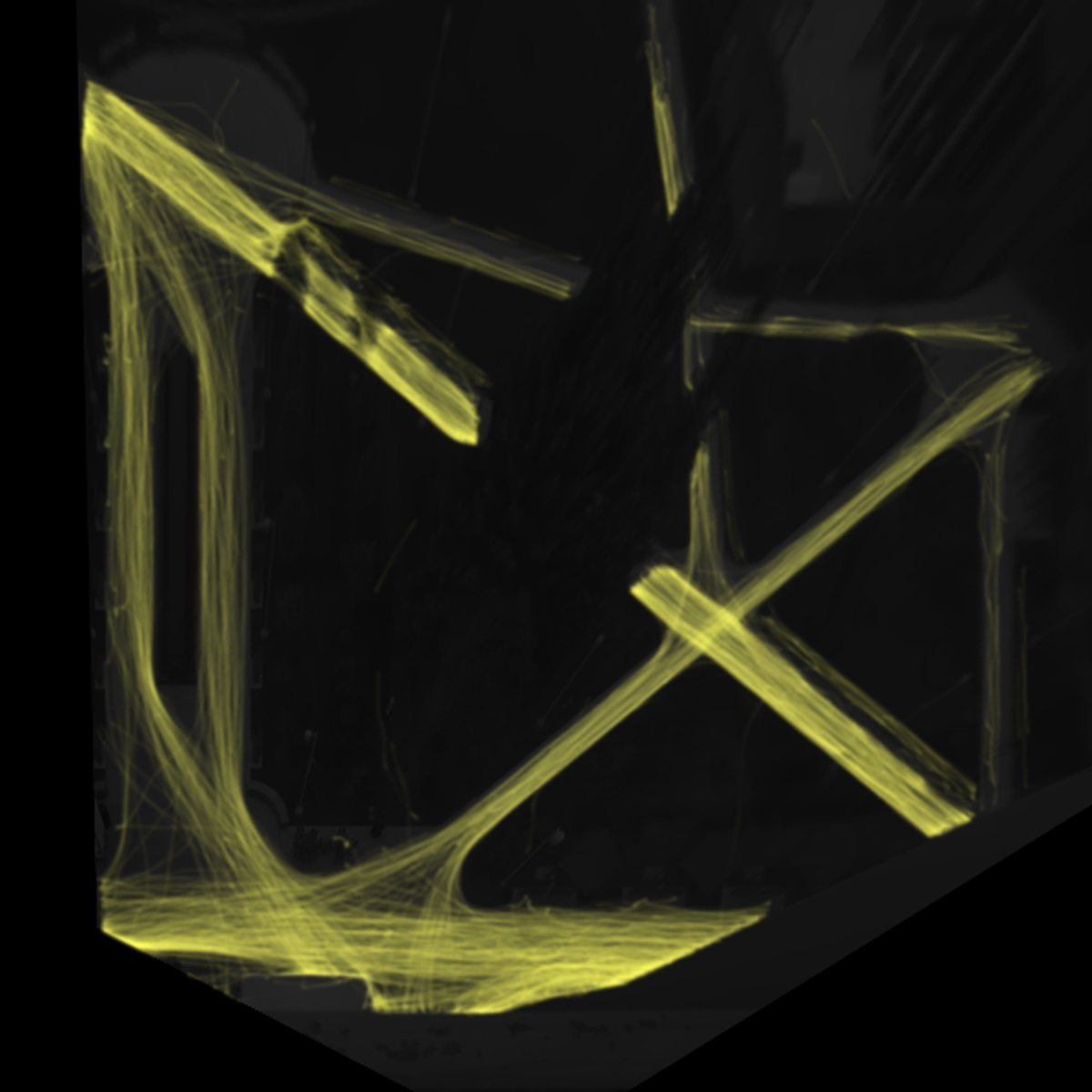}
\caption{Reconstructed pedestrian trajectories from one hour of the
  video recording.}
\label{pedestrian-trajectories}
\end{figure}

To enhance the robustness of tracking, we developed an original,
hybrid algorithm that combined image processing (for object detection
and perspective transformation; implemented with OpenCV) with
real-time online agent-based simulation (for motion {\em prediction}
in a noisy dynamic environment; implemented in Python). Specifically,
each detected object was internally represented as an agent with its
unique position and velocity as agent attributes, and their movements
were simulated internally alongside the image processing. After the
object detection was completed for each video frame, the locations of
the detected objects were compared to those of the simulated agents,
and the objects that were close enough to the existing agents were
``claimed'' by those agents, followed by adjustment of the agents'
positions and velocities according to the positions of the claimed
objects. Unclaimed objects were represented as new agents (which might
be aggregated into other existing agents if their trajectories
converged later). The agents that could not claim any object would
remain moving along a linear projection, either until it re-claimed an
object (in this case the tracking would continue) or until it moved
for a certain period of time without claiming an object or moved out
of the recorded spatial area (in these cases the tracking of the agent
would be terminated and its trajectory would be written out into a
file). Each of the recorded trajectories was smoothed out using a
Gaussian kernel (with standard deviation set to 20 frames $\approx$
2/3 sec.)  applied over time to reduce noise. Figure
\ref{pedestrian-trajectories} shows an example of the reconstructed
pedestrian trajectories.

\begin{figure}
\centering
\includegraphics[width=\columnwidth]{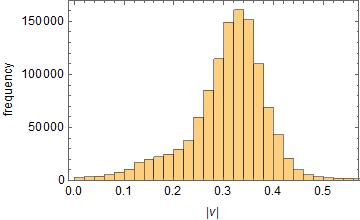}
\caption{Histogram of pedestrians' speeds. In most cases, their speeds
  fell within the range $[0.1, 0.5]$ (unit: pixel per frame).}
\label{pedestrian-histogram}
\end{figure}

\begin{figure}
\centering
\includegraphics[width=\columnwidth]{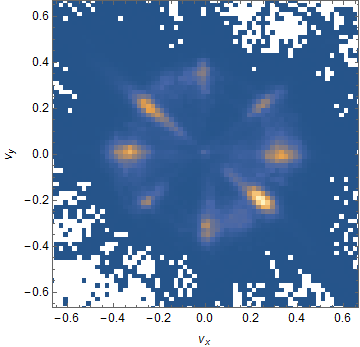}
\caption{Density plot showing the distributions of pedestrians'
  two-dimensional velocities. There were eight clearly established
  directions, reflecting the road patterns seen in
  Fig.~\ref{top-view}.}
\label{pedestrian-velocities}
\end{figure}

Figure \ref{pedestrian-histogram} presents the histogram of the
pedestrians' speeds, which shows that most of the pedestrians moved
with the speed between 0.1 and 0.5 pixels per frame. We therefore
designated the speeds that fell within this range as ``normal
moving'', while the speeds less than 0.1 as ``not moving'', and those
above 0.5 as ``fast moving''. The two moving categories (normal moving
and fast moving) were further classified into eight different
directions, given the eight distinct peaks of velocities observed in
the velocity distribution (Fig.~\ref{pedestrian-velocities}), which
reflected the road patterns seen in Fig.~\ref{top-view}. As a result,
a total of $1+8+8=17$ distinct behavioral states were identified. The
trajectories of pedestrians were labeled with these behavioral states
based on the speed and direction of their movements.

\begin{figure}
\centering
\includegraphics[width=\columnwidth]{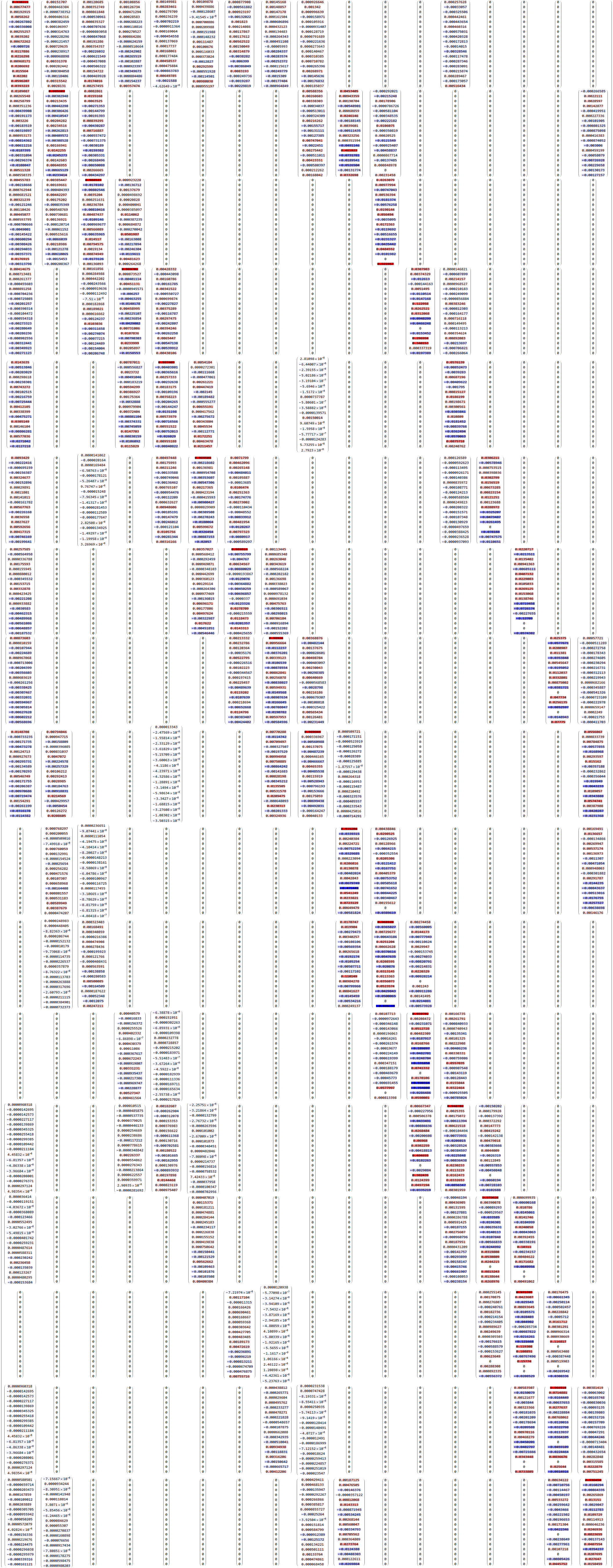}
\caption{State transition tensor $\hat{T}$ obtained from the
  trajectories of pedestrians. The tensor was transposed to improve
  readability. The size of $\hat{T}$ is $17\times 17 \times 18$. The
  format of presentation is the same as in
  Fig.~\ref{termite-tensor}. See Fig.~\ref{enlarged-tensor} for an
  enlarged version.}
\label{pedestrian-tensor}
\end{figure}

\begin{figure*}
\centering
\includegraphics[width=0.9\textwidth]{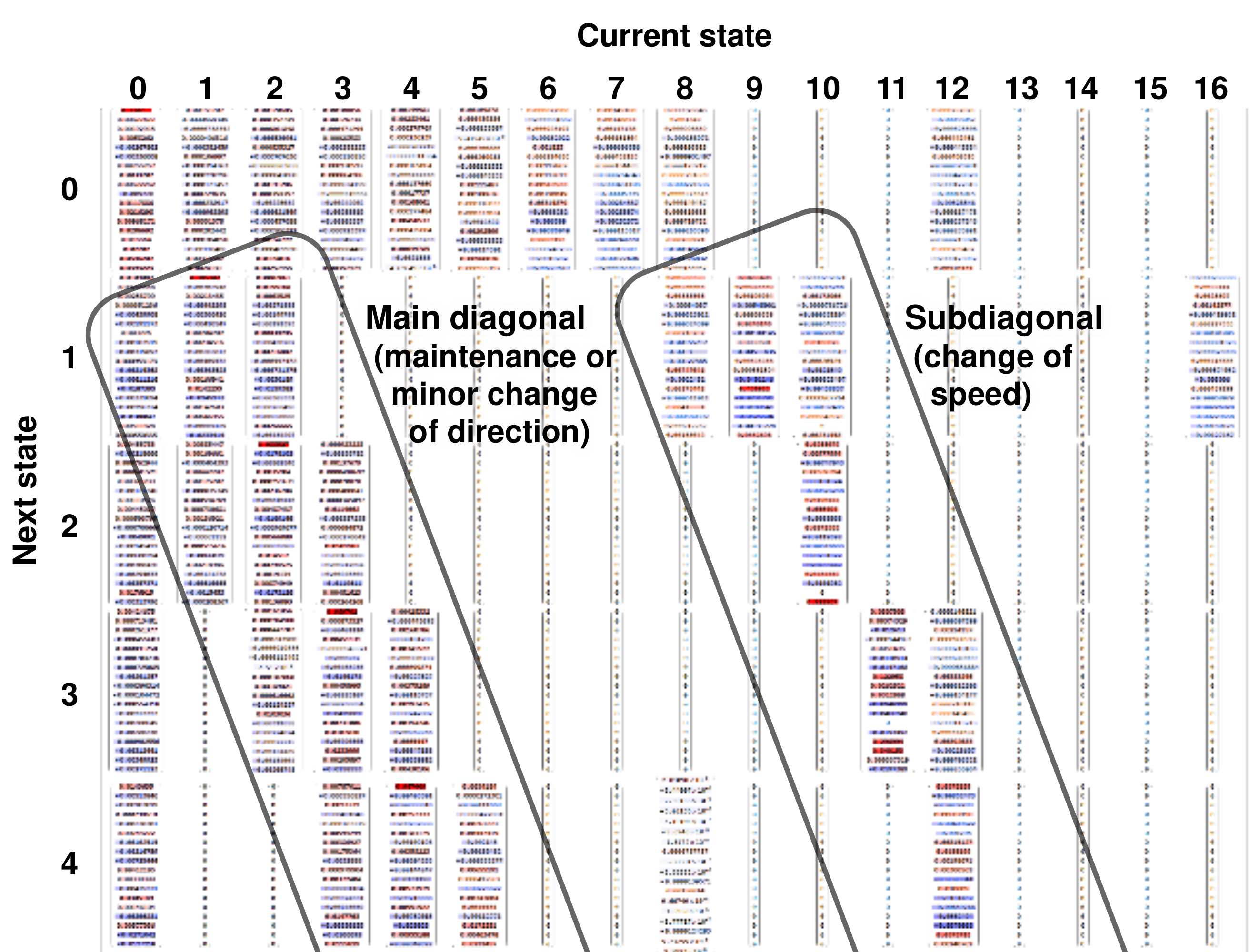}
\caption{Enlarged top part of state transition tensor $\hat{T}$ shown
  in Fig.~\ref{pedestrian-tensor}. Red and blue highlights indicate
  positive and negative coefficients, respectively. Non-zero entries
  typically appear in the main diagonal and subdiagonal parts; the
  former corresponds to the maintenance or minor change of the
  direction of movement, while the latter corresponds to the change of
  speed between normal and fast moving.}
\label{enlarged-tensor}
\end{figure*}

The trajectories labeled with behavioral states were then analyzed by
the same behavioral modeling method as used in the first application,
by counting the numbers of other nearby individuals in each of the 17
states and using the least squares method to estimate the state
transition matrix for each state. Individuals were considered
neighbors if they existed within spatial distance of 30 pixels
($\approx$ 3m) and temporal distance of 15 frames ($\approx$ 0.5 sec.)
from each other. The resulting behavioral transition model was given
as a $17\times 17 \times 18$ tensor (Figs.~\ref{pedestrian-tensor},
\ref{enlarged-tensor}). The majority of transitions were detected in
the main diagonal and subdiagonal parts of the tensor
(Fig.~\ref{enlarged-tensor}). The main diagonal part corresponds to
the maintenance or minor change of the direction of movement, while
the subdiagonal part corresponds to the change of speed between
normal and fast moving. A few notable interactions between individuals
with different states were also detected, such as having neighbors
moving in other directions slows down fast-moving individuals,
etc. Overall, the state interactions were generally much less among
pedestrians than among termites reported in the previous section. This
is because human pedestrians would typically move straight toward an
intended destination. This result demonstrated that the developed
modeling method and tracking system were able to process collective
behaviors in a noisy real-world environment.

\section{Conclusions}

In this paper, we have proposed computational approaches to track the
movements of biological collectives from ordinary video recordings and
to detect and model interactions among individuals with different
behavioral states. Our modeling method was based on the assumption
that the individuals are finite state machines, which facilitated the
identification of behavioral states and the modeling of their state
transitions. We conducted a preliminary evaluation of the proposed
method through two applications, one for termites and the other for
human pedestrians, which both demonstrated the effectiveness of the
proposed method and the developed tracking systems.

Needless to say, there are limitations in the present study. Several
parameters were involved in the proposed method, such as the
spatial/temporal radii used for neighbor detection and the size of the
time window for behavioral state characterization, whose effects on
the results were not fully explored yet. Also, both the termite and
pedestrian examples were observed in highly constrained settings
(i.e., termites were confined in a round Petri dish, and pedestrians
were moving mostly on simple straight walkways). More systematic
evaluations of the effects of parameters and testing the proposed
method with video recordings of collective behaviors in less
constrained open space will be among our future work. We also plan to
conduct critical comparison of our method with other existing
tracking/modeling methods.

We believe that the proposed method has potential for practical
applications. One application area would be to objectively quantify
and compare the strength of local interactions among individuals
within biological collectives across multiple species. Another
application area would be to detect behavioral anomalies, especially
anomalies in interaction patterns, in a crowd of people. Both
inter-species comparison of collective behaviors and anomaly
detection in interaction patterns of organisms are relatively
under-explored research areas. We hope that the proposed method
uniquely sheds light on those aspects of collective dynamics to
further our understanding of complex systems.

\section*{Acknowledgments}

H.S.\ thanks financial support from the US National Science Foundation
(1319152). J.S.T.\ thanks financial support from the Human Frontiers
Science Program (HFSP) (RGP0066/2012).

\bibliographystyle{IEEEtran}
\bibliography{szjt}

\end{document}